\documentclass[preprint,prb,preprintnumbers,amsmath,amssymb]{revtex4}

\usepackage{txfonts}
\usepackage[dvipdfmx]{graphicx}
\usepackage{bm}
\usepackage{times} 
\begin{document}

\title{Ferromagnetic Cluster Glass Phase Embedded in a Paramagnetic and Metallic Host
in Non-uniform Magnetic System CaRu$_{1-\textit{x}}$Sc$_{\textit{x}}$O$_{3}$}

\author{Takafumi D. Yamamoto$^1$$^\ast$, Atsuhiro Kotani$^2$, Hiroshi Nakajima$^2$, \\Ryuji Okazaki$^3$, Hiroki Taniguchi$^1$, Shigeo Mori$^2$, and Ichiro Terasaki$^1$}

\affiliation{$^1$Department of Physics, Nagoya University, Nagoya 464-8602, Japan\\
$^{2}$Department of Materials Science, Graduate School of Engineering, Osaka Prefecture University, Osaka 599-8531, Japan\\
$^{3}$Department of Physics, Faculty of Science and Technology, Tokyo University of Science, Chiba 278-8510, Japan}

\begin{abstract}
We have investigated both static and dynamic magnetic properties of polycrystalline CaRu$_{1-\textit{x}}$Sc$_{\textit{x}}$O$_{3}$
system in order to clarify the role of Sc ions as a disorder for magnetic ordering.
We have observed typical features of a ferromagnetic cluster glass state below around 40 K:
(i) a broad, frequency-dependent peak in the ac magnetic susceptibility,
(ii) a slow relaxation of the magnetization,
and (iii) a continuous increase in the dc magnetic susceptibility in field cooling process.
The composition dependence of characteristic parameters for the cluster glass state suggests that
chemical segregation can hardly explain the clustering mechanism.
We propose a possible picture that the ferromagnetic clusters are distributed uniformly and form the glassy state embedded in
the paramagnetic and metallic host of CaRuO$_{3}$.
\end{abstract}

\maketitle
\newpage
\section{Introduction}
The Ruddlesden-Popper-type ruthenium oxides (Ca, Sr)$_{n+1}$Ru$_{n}$O$_{3n+1}$ have been extensively studied
because of a variety of electronic and magnetic ground states, such as
a spin-triplet superconductor, an antiferromagnetic Mott-insulator, and a ferromagnetic metal\cite{Maeno,NakatsujiJPSJ, Neumeier}.
This richness originates from the fact that
Ru$^{4+}$ ions have more extended d orbitals than 3d transition metal oxides.
Namely, various ground states emerge through two kinds of interplay.
One is the interplay between itinerant and localized nature of 4d electrons.
Strong hybridization between 4d orbitals produces relatively wide 4d bands, being favorable for metallic conduction.
This itinerant character competes with the localized character resulting from the on-site Coulomb repulsion.
The other is the interplay among the charge, spin, orbital, and the lattice degrees of freedom.
Owing to the spatially extended 4d orbitals, Ru$^{4+}$ ions experience the strong crystal field from
six surrounding O$^{2-}$ ions coordinated octahedrally.
As a result, the electronic states of Ru$^{4+}$ are sensitive to the distortion of a RuO$_{6}$ octahedron.
In other words, they are easily modified by changing the lattice degrees of freedom.  
Indeed, the ground state of these ruthenates is tuned by applying a perturbation such as
chemical substitution, pressure, and magnetic field.\cite{NakatsujiPRL,Nakamura,Ikeda,Yoshida}
\par
A nearly cubic perovskite CaRuO$_{3}$ ($n=\infty$) shows a metallic-like conduction and 
a paramagnetic susceptibility obeying the Curie-Weiss law.\cite{Koriyama,CaoRes}
A large negative Curie-Weiss temperature of -140 K could lead to a magnetic transition at low temperature below around 100 K,
but curiously enough, no long-range magnetic ordering has been found down to 1.5 K.\cite{Martinez,Gibb}
Kiyama et al. have found that the characteristics of the specific heat and high-field magnetization can be explained
in the framework of spin-fluctuation theory on itinerant-electron magnetism.\cite{KiyamaC,KiyamaM}
Furthermore, they have observed robust ferromagnetic spin fluctuations in $^{17}$O NMR study.\cite{Yoshimura}
Based on these results, they have suggested that CaRuO$_{3}$ is a nearly ferromagnetic metal.
In contrast, Felner et al. have claimed that this compound is a spin-glass system because the magnetic susceptibility
shows the irreversibility between zero-field cooling and field cooling processes in low magnetic fields.\cite{Felner}
Other groups have discussed from non-Fermi liquid behavior a possibility of a quantum criticality.\cite{CaoQCP,Baran}
In this way, the magnetic ground state of CaRuO$_{3}$ is still under debate.
Nevertheless, there is a consensus that this paramagnetic metal has magnetic instability and
would readily turn into a magnetically ordered state by a perturbation.
\par
Chemical substitutions can be useful to understand the physical properties of such a system.
Many extensive studies on the 3d transition metal substitution for Ru site have suggested that 
a ferromagnetic ordering is induced by a small amount of substitution of both magnetic and non-magnetic ions
such as Cr$^{3+}$, Fe$^{3+}$ and Ti$^{4+}$(Refs. 18-20).
This phenomenon probably reflects the magnetic instability of CaRuO$_{3}$.
Thus, understanding of the ferromagnetic ordering provides an insight into
the magnetic ground state of the paramagnetic metal,
but the nature and origin of the ferromagnetism have not been fully clarified in spite of many efforts so far.
\par
In our previous study,\cite{Previous} we investigated the static magnetic properties
of CaRu$_{1-\textit{x}}$Sc$_{\textit{x}}$O$_{3}$ in the range of 0 $\leq$ \textit{x} $\leq$ 0.20.
We found that all the Sc-substituted samples show a ferromagnetic ordering
with an \textit{x}-independent transition temperature $T_{\rm c} \sim$ 30 K. 
We also found that the Curie-Weiss temperature $\theta_{\rm CW}$ dramatically changes from
-150 K at $\textit{x}$ = 0 to +5 K at $\textit{x} =$ 0.20, in contrast to the $\textit{x}$-independent $T_{\rm c}$.
This inconsistency between $T_{\rm c}$ and $\theta_{\rm CW}$ implies that
CaRu$_{1-\textit{x}}$Sc$_{\textit{x}}$O$_{3}$ is a \textit{non-uniform magnetic system},
i.e., a system consisting of more than one magnetic component.
We proposed a phenomenological two-component model and analyzed the static magnetic properties.
Consequently, we successfully described the static magnetic properties of 
Ca[Ru$^{4+}_{1-2\textit{x}}$Ru$^{5+}_{\textit{x}}$]Sc$^{3+}_{\textit{x}}$O$_{3}$ as
a volume average of a paramagnetic component originating from Ru$^{4+}$ ions and
a ferromagnetic one induced by Ru$^{5+}$ ions which are generated by non-magnetic Sc$^{3+}$ substitution.
\par
The two-component analysis does not necessarily mean a phase segregation,
and a microscopic picture of the ferromagnetic component remains unclear.
In particular, the length scale and the distribution pattern of the ferromagnetic region are open to question.
We note here that the chemical substitution usually introduces a disorder,
and whether a long-range magnetic ordering exists is nontrivial.
In fact, some groups have pointed out a possibility that
CaRu$_{1-\textit{x}}$M$_{\textit{x}}$O$_{3}$ ($M =$ transition metal ions) is an inhomogeneous ferromagnetic system,\cite{Hardy,HeCondens}
but the direct evidence has not yet been found.
In this study, we have investigated the static and dynamic magnetic properties of
CaRu$_{1-\textit{x}}$Sc$_{\textit{x}}$O$_{3}$ (\textit{x} $=$ 0.10, 0.20)
in order to verify the role of Sc ions as a disorder for the magnetic ordering.
The samples used here are the same as used in our previous work.
\section{Experiments}
\begin{figure}[htp]
 \begin{center}
  \includegraphics[scale=0.60,clip]{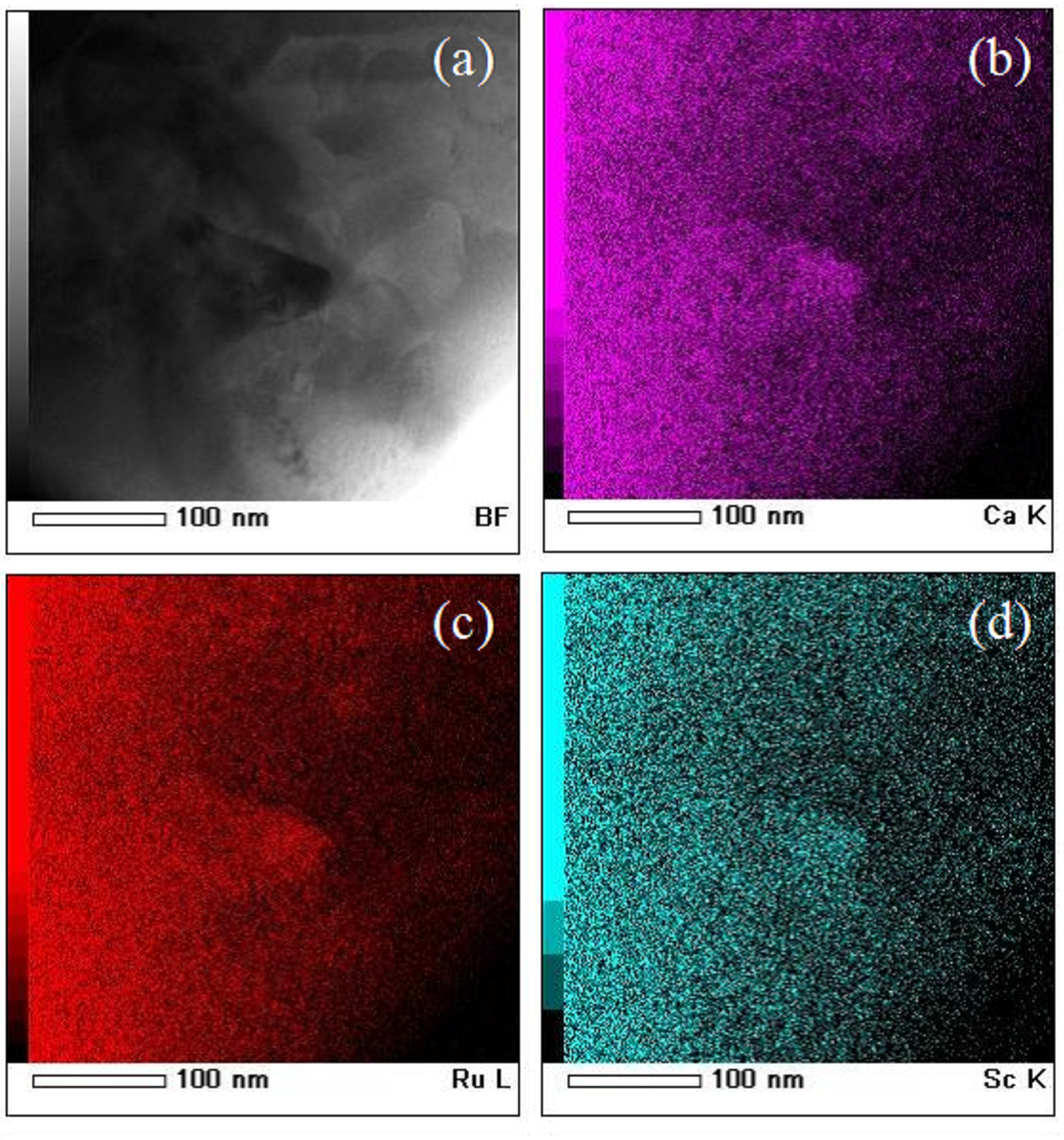}
  \caption{(Color Online) (a) TEM bright-field image of polycrystalline CaRu$_{0.80}$Sc$_{0.20}$O$_{3}$.
(b)-(d) The corresponding energy-dispersive X-ray spectroscopy mapping images for (b) Ca, (c) Ru, and (d) Sc elements.}
\label{fig:EDS}
 \end{center}
\end{figure}

Polycrystalline specimens of CaRu$_{1-\textit{x}}$Sc$_{\textit{x}}$O$_{3}$ ($\textit{x} =$ 0.10, 0.20)
were prepared by the standard solid-state reaction method using high-purity reagents of
$\rm CaCO_3$\ (3N), $\rm RuO_2$\ (3N), and $\rm Sc_2O_3$\ (3N).
A detailed recipe of preparation was described in our preceding paper.\cite{Previous}
X-ray diffraction patterns showed that all the prepared samples crystallize in the GdFeO$_{3}$-type structure
of the space group \textit{Pnma}.
In addition, the diffraction peaks shift systematically in 2$\theta$ and show no split with increasing \textit{x}.
Accordingly, chemical homogeneity of the Sc-substituted samples is likely to be ensured,
though a full width half maximum (FWHM) in 2$\theta$ increased with Sc substitution;
the FWHM of (242) peak changes from 0.20$^{\circ}$ at $\textit{x} =$ 0 to 0.52$^{\circ}$ at $\textit{x} =$ 0.20.
The homogeneity was further examined for CaRu$_{0.80}$Sc$_{0.20}$O$_{3}$ by the energy-dispersive X-ray spectroscopy (EDS)
mapping analysis using  a JEOL JEM-2100F field-emission transmission electron microscope operated at 200 kV. 
Figure \ref{fig:EDS}(a) and (b)-(d) show the TEM bright-field image of the sample and the corresponding EDS mapping images
of Ca, Ru, and Sc elements, respectively. The bright spots stand for the presence of each element.
These images demonstrate that all the elements including Sc are indeed uniformly distributed
throughout the sample with a spatial resolution of 1-2 nm.
\par
The ac magnetic susceptibility measurements were performed between 2 and 60 K in the frequency range from
1 kHz to 100 kHz using a homemade probe. 
The amplitude of an ac magnetic field $h_{\rm ac}$ is about 0.01 Oe.
The static magnetic measurements were carried out using a Quantum Design superconducting quantum interference device magnetometer.
The dc magnetic susceptibility $\chi (= \textit{M/H}$)
in field cooling (FC) and zero-field cooling (ZFC) processes was measured between 2 and 60 K
for an applied dc magnetic field ($H$) of 20 Oe. 
Magnetization ($M$) data were collected between -70 kOe and 70 kOe in the temperature range of 2 $\le T \le$ 80 K.
Magnetic relaxation measurements were performed at 2 K for 24 h in both CaRu$_{0.90}$Sc$_{0.10}$O$_{3}$ and CaRu$_{0.80}$Sc$_{0.20}$O$_{3}$.
In these measurements, the sample was first cooled in ZFC from 200 K down to a desired temperature,
and then a dc magnetic field of 70 kOe was applied for 5 minutes.
After that, the field was switched off and the isothermal remanent magnetization $M_{\rm IRM}(t)$ was recorded
as a function of time $t$.
\section{Results and Discussion}
\begin{figure}[htp]
 \begin{center}
  \includegraphics[scale=0.65,clip]{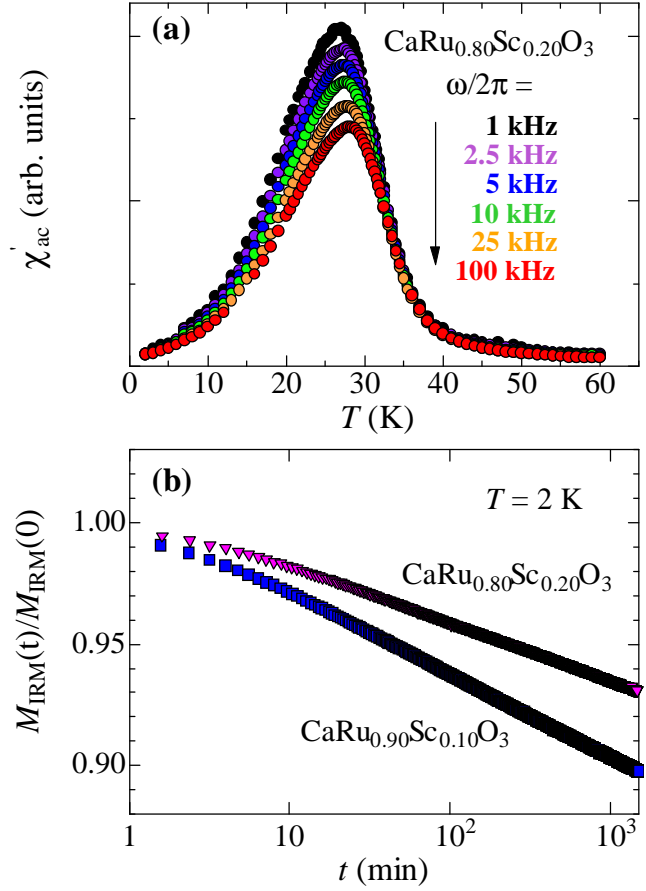}
  \caption{(Color Online) (a) Temperature dependence of the real component of the ac magnetic susceptibility
           $\chi^{'}_{\rm ac}$ for CaRu$_{0.80}$Sc$_{0.20}$O$_{3}$ in the ac magnetic field with various frequencies.
           (b) Time dependence of the isothermal remanent magnetization $M_{\rm IRM}(t)$ of
               CaRu$_{0.90}$Sc$_{0.10}$O$_{3}$ and CaRu$_{0.80}$Sc$_{0.20}$O$_{3}$ measured at 2 K.}
  \label{fig:Dynamic}
 \end{center}
\end{figure}

Figure \ref{fig:Dynamic}(a) shows the temperature dependence of the real component of
the ac magnetic susceptibility $\chi^{'}_{\rm ac}$ for CaRu$_{0.80}$Sc$_{0.20}$O$_{3}$.
$\chi^{'}_{\rm ac}$ exhibits a broad peak at $T_{\rm f}$ for various frequencies $f = \omega/2\pi$.
With increasing $f$, the peak becomes broader and $T_{\rm f}$ shifts to higher temperatures.
The same features are observed in CaRu$_{0.90}$Sc$_{0.10}$O$_{3}$ (not shown).
For conventional ferromagnets, this peak shift is observed in high frequency range from MHz to GHz.\cite{Mydosh}
Consequently, the peak shift observed here suggests that
CaRu$_{1-\textit{x}}$Sc$_{\textit{x}}$O$_{3}$ system does not have a long-range ferromagnetic ordering but rather a magnetic glassy state.
To confirm this, we have measured the time dependence of the isothermal remanent magnetization $M_{\rm IRM}(t)$ at 2 K.
$M_{\rm IRM}(t)$ is found to show a relaxation for each sample as shown in Fig. \ref{fig:Dynamic}(b).
Although this feature is also indicative of the glassiness, the relaxation is too slow
(the decreases of about 9 \% for CaRu$_{0.90}$Sc$_{0.10}$O$_{3}$ and 7 \% for CaRu$_{0.80}$Sc$_{0.20}$O$_{3}$ in a day).
This long relaxation time makes it difficult to distinguish the glassy state from the ferromagnetic ordering
through static magnetic measurements in the Sc-substituted samples.
\par
\begin{figure}[htp]
 \centering
  \includegraphics[scale=0.65,clip]{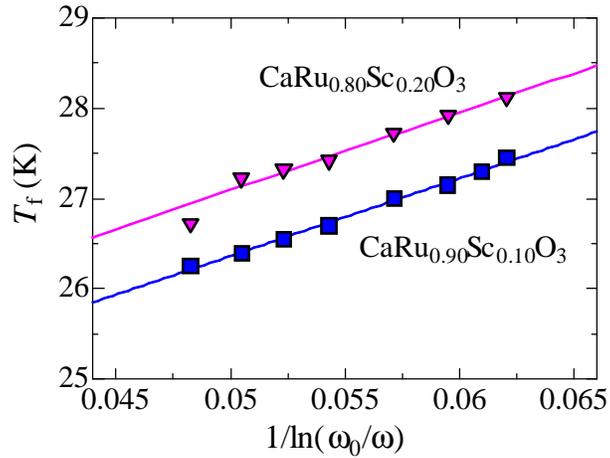}
  \caption{(Color online) The peak temperature $T_{\rm f}$ plotted against 1/ln($\omega_{\rm 0}/\omega$).
            The solid lines represent a fit using the Vogel-Fulcher law (see text).}
  \label{fig:Vogel}
\end{figure}

Here we estimate the frequency-shift rate of $T_{\rm f}$ per decade $\omega$ given by
$\delta T_{\rm f} = (\Delta T_{\rm f}/T_{\rm f})/\Delta \rm{log}_{10} \omega$.
This parameter is useful for characterizing magnetic glassy systems.
The $\delta T_{\rm f}$ value is evaluated to be about 0.022 for both CaRu$_{0.90}$Sc$_{0.10}$O$_{3}$ and CaRu$_{0.80}$Sc$_{0.20}$O$_{3}$.
This value is larger than the values reported for typical spin-glass materials
($\delta T_{\rm f} =$ 0.005 for CuMn and 0.006 for AgMn),\cite{Tholence}
but smaller than those reported for typical superparamagnetic materials
($\delta T_{\rm f} =$ 0.28 for $\alpha$-Ho$_{2}$O$_{3}$(B$_{2}$O$_{3}$)).\cite{Mydosh}
Thus, our system seems to be an intermediate material, i.e., a cluster glass material
which consists of interacting magnetic clusters.
In order to get more information, we have analyzed the $T_{\rm f}$ data by fitting using
the empirical Vogel-Fulcher law given by\cite{Mydosh}
\begin{equation}
\omega = \omega_{0}\ {\rm exp}\left[-\frac{E_{\rm a}}{k_{\rm B}(T_{\rm f}-T_{0})}\right],
\label{eq:Vogel}
\end{equation}
where $k_{\rm B}$ is the Boltzmann constant, $\omega_{0}$ is the characteristic frequency, $E_{\rm a}$ is
the activation energy, and $T_{0}$ is the Vogel-Fulcher temperature. 
A value of $E_{\rm a}$ and $T_{0}$ reflects the size of a cluster and the strength of the intercluster interaction, respectively.
Rewriting Eq. (\ref{eq:Vogel}) as
\begin{equation}
T_{\rm f} = T_{\rm 0}+\frac{E_{\rm a}}{k_{\rm B}}\left[\rm{ln}\left(\frac{\omega_{0}}{\omega}\right)\right]^{-1}
\label{eq:Re-Vogel}
\end{equation}
useful for analyzing the data.
According to this expression, $T_{\rm f}$ is a linear function of $1/\rm{ln}(\omega_{0}/\omega)$.
As shown Fig. \ref{fig:Vogel}, the observed $T_{\rm f}$ data have indeed good linearity. 
The solid lines represent the best fit to the experimental data by Eq. (\ref{eq:Re-Vogel}),
where we fixed $\omega_{0}/2\pi$ to be 10$^{12}$ Hz according to a previous study.\cite{Kawasaki}
From the fitting, we have found that the $E_{\rm a}/k_{\rm B}$ value ($\sim$ 86 K) is unchanged by Sc substitution,
while $T_{0}$ slightly increases from 22.1 K to 22.8 K with increasing \textit{x}.
A non-zero value of $T_{0}$ indicates that the ferromagnetic clusters interact with each other,
from which we can exclude a possibility that our system is superparamagnetic.
Besides, the increase in $T_{0}$ is agreement with the slower relaxation of $M_{\rm IRM}(t)$
in CaRu$_{0.80}$Sc$_{0.20}$O$_{3}$ than that in CaRu$_{0.90}$Sc$_{0.10}$O$_{3}$.
\par
\begin{figure}[htp]
 \centering
  \includegraphics[scale=0.7,clip]{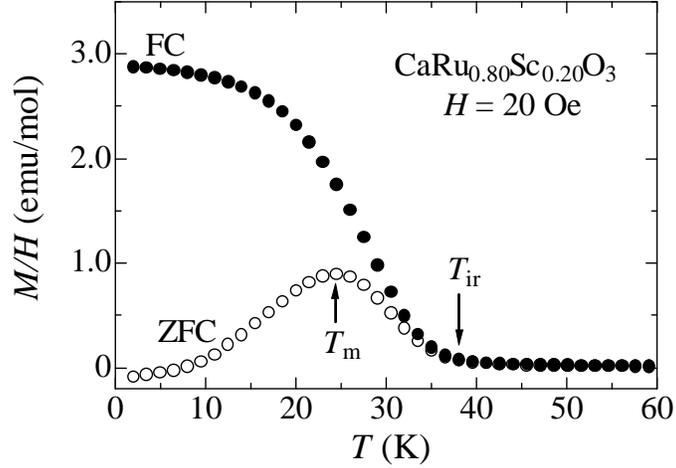}
  \caption{Temperature dependence of the dc magnetic susceptibility $\chi(=M/H)$ for CaRu$_{0.80}$Sc$_{0.20}$O$_{3}$ on field cooling ($\chi_{\rm FC}$)
            and zero-field cooling ($\chi_{\rm ZFC}$) processes in the dc magnetic field of 20 Oe.
		$T_{\rm m}$ and $T_{\rm ir}$ show a peak temperature of $\chi_{\rm ZFC}$ and a bifurcation
		point between $\chi_{\rm ZFC}$ and $\chi_{\rm FC}$, respectively.}
  \label{fig:MT-LH}
\end{figure}

An additional piece of evidence for the cluster glass state is observed in the dc magnetic susceptibility measurements.
Figure \ref{fig:MT-LH} shows the temperature dependence of $M$/$H$ for CaRu$_{0.80}$Sc$_{0.20}$O$_{3}$ on field cooling ($\chi_{\rm FC}$)
and zero-field cooling ($\chi_{\rm ZFC}$) processes in the dc magnetic field of 20 Oe.
There is no difference between $\chi_{\rm FC}$ and $\chi_{\rm ZFC}$ far above 40 K.
With decreasing temperature, $\chi_{\rm ZFC}$ deviates from $\chi_{\rm FC}$ at a bifurcation point $T_{\rm ir}$
($\sim$ 38.0 K) and shows a pronounced maximum at a peak temperature $T_{\rm m}$ ($\sim$ 24.5 K).
An important feature is that $T_{\rm ir} > T_{\rm m}$, which occurs
in ferromagnetic cluster glass systems, not in spin-glass systems.\cite{Nagata}
Another feature is that $\chi_{\rm FC}$ continues to increase below $T_{\rm m}$,
which also occurs in ferromagnetic cluster glass systems.\cite{Li-Ir,Li-Rh}
A spin-glass system would show $T$-independent $\chi_{\rm FC}$ below $T_{\rm m} (= T_{\rm ir})$.\cite{Nagata}
The irreversibility between $\chi_{\rm ZFC}$ and $\chi_{\rm FC}$ below $T_{\rm ir}$ is a sign of
the freezing in spin-glass systems, while the higher $T_{\rm ir}$ than $T_{\rm m}$ in cluster glass systems probably reflects
the formation of the clusters.
Thus, we should consider $T_{\rm m}$ as a freezing temperature of the ferromagnetic clusters.
\par
\begin{figure}[htp]
 \centering
  \includegraphics[scale=0.60,clip]{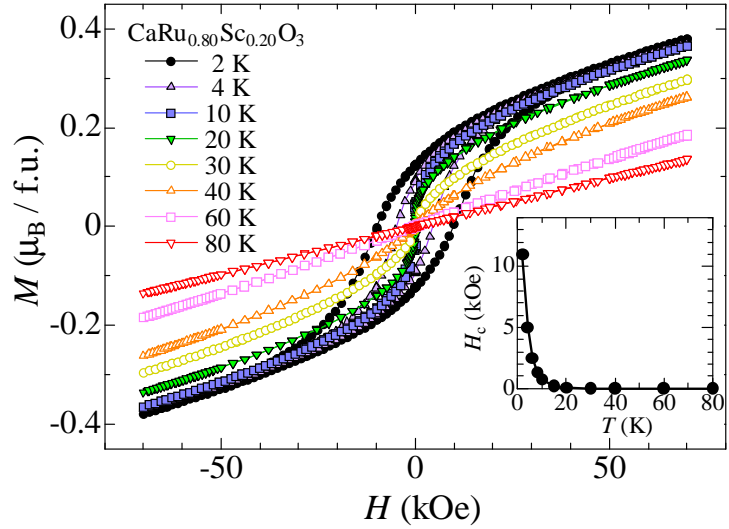}
  \caption{(Color online) Field dependence of the magnetization $M$ in Ca$_{0.80}$Sc$_{0.20}$O$_{3}$ at various temperatures.
           The inset shows the temperature dependence of the coercive force $H_{\rm c}$.}
\label{fig:MH}
\end{figure}

To investigate how the ferromagnetic cluster glass state evolves, we measured the field dependence of magnetization ($M$-$H$ curve).
Figure \ref{fig:MH} shows the $M$-$H$ curves in CaRu$_{0.80}$Sc$_{0.20}$O$_{3}$ measured at various temperatures between 2 and 80 K.
$M$ increases linearly with increasing magnetic field at 80 K, showing that the system is in a paramagnetic state.
With decreasing temperature, the magnetization becomes non-linear significantly below 40 K.
This result suggests that the ferromagnetic clusters begin to form gradually from this temperature,
which is consistent with the rapid increase in $\chi_{\rm FC}$ below $T_{\rm ir}$.
In this case, the increase of $M$ can be explained in terms of the increase of the volume fraction of the clusters.
Since $M$ monotonically increases down to 2 K, the number and/or size of the clusters seems to increase
even after they freeze at $T_{\rm m}$.
The absence of saturation must have resulted from the glassiness.
The inset of Fig. \ref{fig:MH} shows the temperature dependence of the coercive force $H_{\rm c}$,
which is a quantitative measure of magnetic hysteresis.
Magnetic hysteresis loops appear below almost $T_{0}$.
This fact is reasonable because the magnetic interaction between clusters probably causes the magnetic hysteresis.
\par
Next let us discuss a microscopic picture of the ferromagnetic cluster glass state.
The composition dependence of $E_{\rm a}$ and $T_{0}$ suggests that the ferromagnetic clusters with a specific size
increase in number with increasing Sc content, or equivalently, Ru$^{5+}$ ions. 
This fact rules out the simplest possibility for the clustering mechanism that
the ferromagnetic clusters result from domains of densely condensed Ru$^{5+}$ ions driven by
phase segregation/ precipitation, because the cluster size is expected to change in this case.
Here we shall consider the cluster size in relation to the two-component model proposed in our previous report.\cite{Previous}
In this model, the experimentally-observed dc magnetic susceptibility $\chi(\textit{x}, T)$
and magnetization $M(\textit{x}, H)$ of the Sc-substituted samples can be described using the expressions given by
\begin{align}
\chi (\textit{x}, T) &= (1- 2 \textit{x}) \chi_{\rm p}(T) + \textit{x} \chi_{\rm f}(T),\\
M(\textit{x}, H) &= (1- 2 \textit{x})M_{\rm p}(H) + \textit{x}M_{\rm f}(H),
\end{align}
where $\chi_{\rm p}(T)$ and $M_{\rm p}(H)$ are the susceptibility and magnetization of
the paramagnetic component respectively, $\chi_{\rm f}(T)$ and $M_{\rm f}(H)$ are those of the ferromagnetic one.
Note that the ferromagnetic cluster glass state should lead to the ferromagnetic component.
The scaling of $M_{\rm f}$ to \textit{x} implies that the clusters do not overlap with each other
despite increase in the number of clusters with \textit{x}.
To satisfy this condition up to \textit{x} $=$ 0.20, the length scale of the ferromagnetic clusters should be comparable with the Ru-Ru distance.
In this context, a minimum cluster will be six Ru$^{4+}$ ions surrounding one Ru$^{5+}$ ion.
Accordingly, we propose a possible picture that the small ferromagnetic clusters are distributed uniformly
at a nano-meter scale and form the glassy state embedded in the paramagnetic and metallic host of CaRuO$_{3}$.
For this picture, the clusters are proportional to \textit{x} in number.
It is, however, yet to be explored whether such a microscopic clusters really exist or not.
\par
Finally, we mention a point of similarity between non-magnetic ion substitutions.
We notice that CaRu$_{1-\textit{x}}$Ti$_{\textit{x}}$O$_{3}$ system shows static magnetic properties
similar to the Sc-substituted samples:
the \textit{x}-independent onset of the dc magnetic susceptibility and the absence of saturation in the magnetization.\cite{Hardy}
This fact implies that the non-magnetic ion substitutions induce a common magnetic state, i.e., the paramagnetic
phase originating from CaRuO$_{3}$ and the ferromagnetic cluster glass phase driven by the substitution.
For clarifying the common features, a comprehensive study of non-magnetic ion substitution effects is necessary.
\section{Summary}
We have investigated the static and dynamic magnetic properties of polycrystalline CaRu$_{1-\textit{x}}$Sc$_{\textit{x}}$O$_{3}$ ($\textit{x} =$ 0.10, 0.20).
We find the signature of a magnetic glass state:
the frequency dependence of the ac magnetic susceptibility in the range of 1 k $\le \omega$/2$\pi \le$ 100 kHz 
and the slow relaxation of the isothermal remanent magnetization.
The frequency shift rate of $T_{\rm f}$ is evaluated to be 0.022, which corresponds to that of cluster glass systems.
Furthermore, the dc magnetic susceptibility shows characteristics similar to those observed in cluster glass systems. 
These results have verified the presence of the ferromagnetic cluster glass phase.
We also find that the magnetization shows the non-linear field dependence below 40 K.
This result suggests that the ferromagnetic clusters begin to form from this temperature
and freeze at $T_{\rm m}(\sim$ 24.5 K).
The \textit{x} dependence of the characteristic parameters, $E_{\rm a}$ and $T_{0}$, suggests that
the ferromagnetic clusters increase in number while keeping their size unchanged.
We have proposed a possible picture to be examined with further microscopic experiments.

\section*{Acknowledgements}
\addcontentsline{toc}{section}{ACKNOWLEDGMENTS}
This work was partially supported by a Grant-in-Aid for Scientific Research, MEXT, Japan (Nos. 25610091, 26247060).
One of the authors (T. D. Yamamoto) was supported by Program for Leading Graduate Schools "Integrative Graduate
Education and Research in Green Natural Sciences", MEXT, Japan.
\newpage

\newpage

\end{document}